\documentclass[prl,twocolumn]{revtex4}
\usepackage{graphicx}
\usepackage{amsmath}

\begin{document}

\title{Theory for the dynamics of glassy mixtures with particle size swaps}

\author{Grzegorz Szamel}
\affiliation{Department of Chemistry, 
Colorado State University, Fort Collins, CO 80523}

\date{\today}

\begin{abstract}
We present a theory for the dynamics of binary mixtures with particle size 
swaps. The general structure of the theory shows that, 
in accordance with physical intuition, particle size swaps open 
up an additional channel for the relaxation of density fluctuations. Thus, allowing
particle size swaps speeds up the dynamics. To make explicit predictions, we use 
a factorization approximation similar to that employed
in the mode-coupling theory of glassy dynamics. We calculate an approximate 
dynamic glass transition phase diagram for an equimolar binary hard sphere mixture. 
We find that in the presence of particle size swaps,  
with increasing ratio of the hard sphere diameters
the dynamic glass transition line moves towards higher volume
fractions, up to the ratio of the diameters approximately equal to 
1.2, and then saturates. We comment on the implications of our findings 
for the theoretical description of the glass transition.  
\end{abstract}

\maketitle 

\textit{Introduction. --}
Until recently, 
computer simulation studies of supercooled fluids suffered from the inability to
equilibrate model glass-forming systems at temperatures close to those corresponding 
to the laboratory glass transition temperature \cite{BerthierBiroliRMP}. 
Advances in the so-called swap 
dynamics computer simulation algorithms allowed researchers to overcome 
this restriction for a class of glass-forming systems \cite{BerthierCNO,NinarelloPRX}. 
Swap dynamics algorithms add Monte Carlo moves in which exchanges of the diameters of 
two different particles are attempted to standard local Monte Carlo or 
Molecular Dynamics simulations. 
It has been found that the slow down of the swap 
dynamics algorithms with increasing density and/or decreasing temperature is much
less drastic than that of the 
standard local Monte Carlo or Molecular Dynamics algorithms. Since swap dynamics 
generates the same equilibrium ensemble as either local Monte Carlo or 
Molecular Dynamics, more gradual slowing down makes possible equilibration of certain
model systems at temperatures equal to or below those corresponding to the
laboratory glass transition temperature, which enables studies of equilibrium 
properties of deeply supercooled fluids and glassy solids 
\cite{BerthierPNAS}. 

The advances in the swap dynamics algorithms lead to interesting theoretical
questions. First, why is swap dynamics so much faster than ``normal'' dynamics?
Second, does the swap dynamics speed-up have implications for the description
of supercooled fluids dynamics and the glass transition? In particular, how
can the difference between the dynamics without and with swaps be reconciled with
the so-called Random First Order Transition (RFOT) 
framework, which connects slowing down upon approaching the
glass transition with changes of static quantities, the configurational entropy and the 
static point-to-set correlation length  
\cite{BerthierBiroliRMP,BouchaudBiroli,LubchenkoWolynes}. Here we briefly 
comment on the former question; we will return to the latter one at the end of this
Letter. 

While intuitively it seems plausible that exchanges of particles'
radii result in significant changes of local neighborhoods, which should speed up the 
relaxation of density fluctuations, a theoretical description of this
speed-up is lacking. In a recent preprint Brito \textit{et al.} \cite{BritoLernerWyart} 
asserted, on the basis of general arguments applicable only to systems with 
continuous polydispersity, that allowing particle size swaps 
results in the decrease of the 
onset temperature for glassy behavior. However, their approach does not lead to a 
specific quantitative prediction for this change. Somewhat earlier, Ikeda \textit{et al.} 
\cite{IkedaZamponiIkeda} used the correspondence between the so-called (avoided)  
dynamic glass transition observed in simulations \cite{KobLH} 
and predicted by approximate theories \cite{Goetzebook}, and the dynamic
transition predicted by %the static 
replica theory \cite{ParisiZamponi}.  These two transitions coincide in the only exactly 
solvable glassy particle-based model, the infinitely dimensional model of spherically 
symmetric particles \cite{MaimbourgPRL,largedCKPUZAR}. Ikeda \textit{et al.} argued that 
the presence of the exchanges of particles' diameters implies a more general structure 
of the \textit{Ansatz} for the inter-replica correlation for a simple 
mean-field-like glass-forming model, the binary 
Mari-Kurchan model \cite{MariKurchan,CJPZPNAS2014}.
They showed that the new \textit{Ansatz} allows one to distinguish between 
dynamic transitions without and with exchanges of particles' diameters.

In this Letter we present a \textit{dynamic} theory for the acceleration due
to the particle size swaps. First, on the basis of the general structure of the theory,
we argue that particle size swaps open an additional relaxation channel, 
which speeds up the dynamics. 
Then, to make explicit predictions, we 
use a factorization approximation to evaluate irreducible memory functions 
\cite{Goetzebook}. We calculate 
an approximate dynamic glass transition phase diagram for an equimolar binary 
hard sphere mixture. In the presence of particle size swaps, with increasing 
ratio of the diameters the dynamic transition 
shifts towards higher volume fractions. The shift 
saturates at about  4\% at the diameter ratio of approximately 1.2. 

\textit{Model: binary mixture with particle size swaps. --} We consider a binary mixture, which is 
the simplest model that allows one to investigate the influence of the particle size 
swaps on the dynamics. In a recent study of swap algorithms \cite{NinarelloPRX} 
it was found that particle size exchanges in systems with a continuous polydispersity 
result in the largest speed-up of the dynamics. Continuous polydispersity has
some theoretical advantages \cite{comment}, but there are no 
approximate expressions for equilibrium pair correlation functions for systems with
continuous polydispersities, which makes explicit calculations difficult. 

We consider a binary mixture consisting of $N$ particles in volume $V$. Particles
can be of type $A$ or $B$, which differ by size. Since any particle 
can change its type, the state of particle $i$ is determined by its position,
$\mathbf{r}_i$, and type indicated by a binary variable $\sigma_i$,
with $\sigma_i=1$ corresponding to $A$ and $\sigma_i=-1$ corresponding to $B$. 
The composition of the system is specified by the difference of the chemical
potentials of particles of type $A$ and $B$, $\Delta \mu$. We note that the
composition depends also on the number density $n=N/V$ and the temperature $T$
of the system. We will assume that these three parameters result in 
concentrations $x_A$ and $x_B$, and that these concentration are constant while
the density or the temperature of the system varies. In practical calculations 
we will restrict ourselves to equimolar mixtures (for which some formulas simplify).
We assume that the ``normal'' dynamics of the system is Brownian, \textit{i.e.}, that 
each particle moves under the combined influence of thermal noise and interparticle 
forces. The forces are derived from a spherically symmetric potential, which depends 
on the particle type, $V_{\sigma_i \sigma_j}(r_{ij})$, where 
$r_{ij}=|\mathbf{r}_i-\mathbf{r}_j|$ is the distance between particles $i$ and $j$.
In addition to the Brownian motion in space, the particles can change their type (size). 
We assume that each particle can change its type independently of the type changes
of other particles. This is analogous to the single-spin-flip dynamics of spin
systems. In contrast, in practical computational applications one typically changes
the types/sizes of two particles in such a way that the number of particles of each
size is conserved, which is analogous to the so-called Kawasaki dynamics of spin
systems. The latter procedure is convenient because it allows one to 
maintain easily a specific composition of the system. We believe that replacing 
the latter procedure by single particle size swaps is a relatively mild change 
and that both procedures lead to qualitatively similar results. 

The above described model corresponds to the following equation of motion 
for the $N$-particle distribution,
$P_N(\mathbf{r}_1,\sigma_1, ..., \mathbf{r}_N,\sigma_N;t)$, abbreviated below as $P_N$,
\begin{eqnarray}\label{Peom}
\partial_t P_N =
\Omega_{\text{sw}}
P_N \equiv \left(\Omega + \delta \Omega_{\text{sw}}\right) P_N.
\end{eqnarray}
Here the evolution operator $\Omega_{\text{sw}}$ consists of two parts
describing two relaxation channels, the part 
describing Brownian motion of particles,
\begin{eqnarray}\label{Omega}
\Omega = D_0 \sum_i \partial_{\mathbf{r}_i}\cdot
\left(\partial_{\mathbf{r}_i}-\beta\mathbf{F}_i\right)
\end{eqnarray}
and the part describing particle size swaps,
\begin{eqnarray}\label{deltaOmegasw}
\delta\Omega_{\text{sw}} = 
- \tau_{\text{sw}}^{-1} \sum_i \left(1-S_i\right) w_i .
\end{eqnarray}
In Eq. (\ref{Omega}) $D_0$ is the diffusion coefficient of an isolated particle,
$D_0=k_BT/\xi_0$, with $\xi_0$ being the friction coefficient of an isolated particle,
$\beta=1/k_B T$, and 
$\mathbf{F}_i$ is the total force on particle $i$, 
$\mathbf{F}_i=\sum_{j\neq i} \mathbf{F}_{ij} = - \sum_{j\neq i} \partial_{\mathbf{r}_i}
V_{\sigma_i\sigma_j}(r_{ij})$. In Eq. (\ref{deltaOmegasw})
$\tau_{\text{sw}}^{-1}$ is the rate of attempted particle size swaps, $S_i$ is the swap 
operator, $S_i \sigma_i = -\sigma_i$, and $w_i$ is the factor ensuring that the 
detailed balance condition, $\left(1-S_i\right)w_i P_N^{eq} = 0$, is satisfied,
with $P_N^{eq}$ being the equilibrium distribution,
$P_N^{eq}\propto \exp\left(-\beta\sum_{i\neq j} V_{\sigma_i\sigma_j}(r_{ij}) + 
\sum_i \frac{\beta}{2}\Delta\mu\sigma_i \right)$. 
The factor $w_i$ depends on the way particle size swaps are attempted. In practical
applications one typically uses Metropolis criterion for accepting attempted swaps. 
It should be emphasized that while the interactions influencing particles' motion
in space are pairwise-additive, the factor $w_i$ typically is not and it depends on
the whole neighborhood of particle $i$. 

The basic object of our theory are the density correlation functions,
\begin{eqnarray}\label{Fabdef}
F_{\alpha\beta}(q;t)  
= \left<n_\alpha(\mathbf{q})e^{\Omega_{\text{sw}}t} 
n_\beta(-\mathbf{q})\right>,
\end{eqnarray}
where $n_{\alpha}$, $\alpha=A,B$, 
are the Fourier transforms of the normalized microscopic densities
of particles of type $\alpha$,
\begin{equation}\label{nabdef}
n_{\alpha}(\mathbf{q}) 
=  \frac{1}{\sqrt{N}} 
\sum_i \frac{1+\sigma_i\left(\delta_{\alpha A}-\delta_{\alpha B}\right)}{2} 
e^{-i\mathbf{q}\cdot\mathbf{r}_i}.
\end{equation}
In Eq. (\ref{Fabdef}) and in the following equations the standard conventions apply: 
$\left<\dots\right>$ denotes the semi-grand canonical ensemble average
over $P_N^{eq}$, the equilibrium probability distribution
stands to the right of the quantity being averaged, and all operators
act on it as well as on everything else.

We should emphasize that, in contrast to the approach of Ikeda \textit{et al.},
in our theory the functions that characterize the dynamics and whose non-zero
long-time limits signal the 
dynamic glass transition are the same for systems evolving without 
and with swap dynamics.

\textit{General theory. --} We use the standard 
projector operator procedure to derive the general structure of the theory
for the dynamics with particle size swaps. First, we define a projection
operator on the density subspace, $\mathcal{P}$
\begin{eqnarray}\label{projdef}
\mathcal{P} = \dots \sum_{\alpha\beta} \left. n_\alpha(-\mathbf{q})\right>
S^{-1}_{\alpha\beta}(q) \left< n_\beta(\mathbf{q}) \right. \dots,
\end{eqnarray}
and the orthogonal projection, $\mathcal{Q}$,
\begin{eqnarray}\label{oprojdef}
\mathcal{Q} = \mathcal{I} - \mathcal{P} \equiv 
\mathcal{I} - \dots \sum_{\alpha\beta} \left. n_\alpha(-\mathbf{q})\right>
S^{-1}_{\alpha\beta}(q) \left< n_\beta(\mathbf{q}) \right. \dots .
\end{eqnarray}
In Eqs. (\ref{projdef}-\ref{oprojdef}) $S_{\alpha\beta}(q)$ denote 
the partial structure factors, 
$S_{\alpha\beta}(q)\equiv F_{\alpha\beta}(q;t=0)$. Next, using
projection operator identities \cite{Goetzebook,CHess,Suppl}
we express the Laplace transforms of the time-derivatives of the density correlation
functions in terms of the \textit{reducible} memory functions,
\begin{eqnarray}\label{FMred}
\lefteqn{ zF_{\alpha\beta}(q;z) - S_{\alpha\beta}(q) = }
\nonumber \\ &&  - \sum_{\gamma\delta} 
\left(O_{\alpha\gamma}(q) - M_{\alpha\gamma}^{\text{red}}(q;z)\right)
S^{-1}_{\gamma\delta}(q)
F_{\delta\beta}(q;z). 
\end{eqnarray}
In Eq. (\ref{FMred}) $O$ is the frequency matrix, 
$O_{\alpha\beta}(q) 
= -\left<n_\alpha(\mathbf{q})\Omega_{\text{sw}}n_\beta(-\mathbf{q})\right>$ and
$M_{\alpha\gamma}^{\text{red}}(q;z)$ 
is the matrix of reducible memory functions. 
In the present case, with two different relaxation channels, it is convenient
to express both matrices in terms of 3-dimensional vectors 
$\mathsf{v}_\alpha$, $\alpha=A,B$, 
and 3x3 matrices $\mathsf{O}$ and $\mathsf{M}^{\text{red}}$,
\begin{eqnarray}\label{O33}
O_{\alpha\beta}(q) &=& \mathsf{v}_{\alpha}^{\text{T}}
\mathsf{O}(q) \mathsf{v}_{\beta},
\\ \label{M33}
M_{\alpha\beta}^{\text{red}}(q;z) 
&=& \mathsf{v}_{\alpha}^{\text{T}}\mathsf{M}^{\text{red}}(q;z) 
\mathsf{v}_{\beta}.
\end{eqnarray}
Here $\mathsf{v}_{A}^{\text{T}}=(1,0,1)$, $\mathsf{v}_{B}^{\text{T}}=(0,1,-1)$, 
$\mathsf{O}_{11}=D_0q^2 x_A$, $\mathsf{O}_{22}=D_0q^2 x_B$, 
$\mathsf{O}_{33}=(1/2N\tau_{\text{sw}})\left<\sum_i w_i\right>$, 
$\mathsf{O}_{ab}=0$ for $a\neq b$, and
the matrix $\mathsf{M}^{\text{red}}$ reads 
\begin{eqnarray}\label{M33red}
\mathsf{M}_{ab}^{\text{red}}(q;z) = 
\left<\pi_a(\mathbf{q}) \left(z-\mathcal{Q}\Omega_{\text{sw}}\mathcal{Q}\right)^{-1}
\pi_b(-\mathbf{q})\right>,
\end{eqnarray}
where 
\begin{eqnarray}\label{pi12}
\pi_{1,2}(\mathbf{q}) \!\! &=& \!\! 
\frac{D_0}{\sqrt{N}}
\mathcal{Q}\sum_i i\mathbf{q}\cdot\left( i\mathbf{q} - \beta\mathbf{F}_i\right)
\frac{1\pm\sigma_i}{2}e^{-i\mathbf{q}\cdot\mathbf{r}_i},
\\ \label{pi3}
\pi_3(\mathbf{q}) \!\! &=& \!\! - \frac{1}{\sqrt{N}\tau_{\text{sw}}} \mathcal{Q} 
\sum_i w_i \sigma_i e^{-i\mathbf{q}\cdot\mathbf{r}_i}.
\end{eqnarray}

As argued by Cichocki and Hess \cite{CHess} and later, more generally, by 
Kawasaki \cite{Kawasaki}, for systems with stochastic dynamics, memory functions
analogous to $\mathsf{M}^{\text{red}}$ can be reduced further by introducing
the so-called irreducible evolution operator \cite{GSdiagram}.

We define the irreducible evolution operator $\Omega_{\text{sw}}^{\text{irr}}$ 
as follows \cite{Suppl}
\begin{eqnarray}\label{Omegaswirr}
&& \Omega_{\text{sw}}^{\text{irr}} = \mathcal{Q}\Omega_{\text{sw}}\mathcal{Q} 
\\ \nonumber  
&& - \sum_{\alpha\beta} \left. \mathcal{Q} \Omega n_\alpha(-\mathbf{q})\right>
\left<n_\alpha(\mathbf{q})\Omega n_\beta(-\mathbf{q})\right>^{-1}
\left<n_\beta(\mathbf{q})\Omega
\mathcal{Q}\right.  
\\ \nonumber 
&& - \left. \mathcal{Q} \delta\Omega_{\text{sw}}\sigma(-\mathbf{q})\right>
\left<\sigma(\mathbf{q})\delta\Omega_{\text{sw}}\sigma(-\mathbf{q})\right>^{-1}
\left<\sigma(\mathbf{q})\delta\Omega_{\text{sw}}
\mathcal{Q}\right. ,
\end{eqnarray}
where $\sigma(\mathbf{q})$ is the microscopic composition field,
\begin{equation}\label{sigmadef}
\sigma(\mathbf{q})= n_A(\mathbf{q})-n_B(\mathbf{q})
= \frac{1}{\sqrt{N}}
\sum_i \sigma_i e^{-i\mathbf{q}\cdot\mathbf{r}_i}.
\end{equation}
We note that reducible parts of $\mathcal{Q}\Omega_{\text{sw}}\mathcal{Q}$
are removed separately for the two relaxation channels in our system.

Using definition (\ref{Omegaswirr}) we can express $\mathsf{M}^{\text{red}}$ in
terms of matrix $\mathsf{M}^{\text{irr}}$ whose elements are functions 
evolving with the irreducible evolution operator,
\begin{eqnarray}\label{M33redirr}
\mathsf{M}^{\text{red}}(q;z) =
\mathsf{M}^{\text{irr}}(q;z)  -
\mathsf{M}^{\text{irr}}(q;z) \mathsf{O}^{-1}(q)
\mathsf{M}^{\text{red}}(q;z),
\end{eqnarray}
where 
\begin{eqnarray}\label{M33irr}
\mathsf{M}_{ab}^{\text{irr}}(q;z) = 
\left<\pi_a(\mathbf{q}) \left(z-\Omega_{\text{sw}}^{\text{irr}}\right)^{-1}
\pi_b(-\mathbf{q})\right>.
\end{eqnarray}
The combination of Eqs. (\ref{FMred}-\ref{M33}) and (\ref{M33redirr}-\ref{M33irr})
defines the general structure of our theory. To appreciate its
meaning it is instructive to consider an approximation that neglects the 
time-delayed coupling between ``normal'' dynamics and particle size swaps. To this end
we set $\mathsf{M}_{ab}^{\text{irr}}(q;z)=0$ 
for $a=1,2$ and $b=3$, and $a=3$ and $b=1,2$.
One can show \cite{Suppl} that in this case the right-hand-side of Eq. (\ref{FMred}) 
becomes a sum of two independent terms (implying two \textit{parallel} 
relaxation channels), the first one originating from  ``normal'' dynamics and the 
second one due to particle size swaps. 
If the relaxation rate due to the first term becomes
very small, the presence of the second channel can dramatically speed up the dynamics.
We should remember, however, that particle size swaps alone cannot equilibrate 
the system. 

\textit{Mode-coupling-like approximation. --}
To make explicit predictions we need to calculate the elements of 
matrix $\mathsf{M}^{\text{irr}}$. To this end we follow  the spirit of the 
mode-coupling theory \cite{Goetzebook}, which is one of the most-successful but also
most-criticized theories for glassy dynamics in three dimensions. Specifically, we use 
the sequence of three approximations \cite{Goetzebook,SL}: we project
functions $\pi_a$ onto the subspace spanned by the parts of density
products orthogonal to the one-particle densities, factorize four-point dynamic 
correlation functions while replacing the irreducible evolution operator by the original 
un-projected evolution operator, factorize four-point static 
correlation functions, and use some additional approximations that
amount to neglecting higher-order correlation functions in the expressions for the 
so-called vertices \cite{Suppl}. In this way we obtain the following approximate 
expressions for the matrix elements of $\mathsf{M}^{\text{irr}}$:
\begin{eqnarray}\label{mct1}  
&& \mathsf{M}^{\text{irr}}_{ab}(\mathbf{q}; t) \approx
\frac{1}{2} \sum_{\alpha,...,\theta}\sum_{\mathbf{k}_1,\mathbf{k}_2}
\left<\pi_a(\mathbf{q}) 
n_\alpha(-\mathbf{k}_1)n_\beta(-\mathbf{k}_2)\right> 
\nonumber \\ && \times 
S_{\alpha\gamma}^{-1}(\mathbf{k}_1)
S_{\beta\delta}^{-1}(\mathbf{k}_2)
F_{\gamma\epsilon}(\mathbf{k}_1;t)
F_{\delta\zeta}(\mathbf{k}_2;t)
\nonumber \\ && \times 
S_{\epsilon\eta}^{-1}(\mathbf{k}_1)
S_{\zeta\theta}^{-1}(\mathbf{k}_2)
\left< n_\eta(\mathbf{k}_1)n_\theta(\mathbf{k}_2)\pi_b(-\mathbf{q})
\right>.
\end{eqnarray}
The vertices originating from Brownian dynamics part of the evolution 
operator have the same form as in the standard mode-coupling theory for 
binary mixtures \cite{Goetzebook,Suppl}. 
For \textit{an equimolar binary mixture} the new vertex, which originates from the particle 
size swaps, reads 
\begin{eqnarray}\label{vertex3}
&& \left< n_\eta(\mathbf{k}_1)n_\theta(\mathbf{k}_2)\pi_3(-\mathbf{q})
\right> 
= - \delta_{\mathbf{k}_1+\mathbf{k}_2,\mathbf{q}}
\frac{\left< \sum_l w_l \right>}{2 N^{3/2}\tau_{\text{sw}}}\sum_{\mu} 
\nonumber \\ && 
\left[\sum_\lambda \left(
S_{\eta\lambda}(k_1)\delta_{\theta\mu} + S_{\theta\lambda}(k_2)\delta_{\eta\mu}\right)
-4 S_{\eta\mu}(k_1)S_{\theta\mu}(k_2)\right]
\nonumber \\ && \times 
\left(\delta_{\mu A}-\delta_{\mu B}\right).
\end{eqnarray}
For non-equimolar mixtures there are additional terms in the new vertex,
which will be reported elsewhere. 
We note that to derive expression (\ref{vertex3}) we factored out the
average $\left< \sum_i w_i \right>$ \cite{Suppl}. This somewhat technical step results 
in the dynamic glass transition being independent of the detailed form of $w_i$. 

The exact memory function representation of the density correlation function,
Eq. (\ref{FMred}) together with Eqs. (\ref{O33}-\ref{M33}, \ref{M33redirr}),
combined with 
the approximate form of $\mathsf{M}^{\text{irr}}$, Eqs. (\ref{mct1}-\ref{vertex3}), 
constitute our mode-coupling-like 
theory for the dynamics of a binary mixture with particle 
size swaps. To get the explicit predictions for the time-dependence one has to solve
these equations numerically.

\textit{Dynamic glass transition. --}
To find the location of the dynamic glass transition we assume that 
as the density increases and/or the temperature decreases density correlation 
functions develop plateaus and at the transition these plateaus do not decay. Thus,
at the transition non-zero long-time limits
$F_{\alpha\beta}(q;\infty)$ appear discontinuously. 
This assumption allows us to derive from our theory the following self-consistent
equation for $F_{\alpha\beta}(q;\infty)$,
\begin{eqnarray}\label{dgt}
\lefteqn{ F_{\alpha\beta}(q;\infty) - S_{\alpha\beta}(q) = }
\\ \nonumber && 
- \sum_{\gamma\delta}
\mathsf{v}^{\text{T}}_\alpha \mathsf{O}(q) 
\left[\mathsf{M}^{\text{irr}}(q,\infty)\right]^{-1} \mathsf{O}(q) \mathsf{v}_\gamma
S_{\gamma\delta}^{-1}(q) F_{\delta\beta}(q;\infty).
\end{eqnarray}
where $\mathsf{M}^{\text{irr}}(q,\infty)$ is given by Eq. (\ref{mct1}) 
with non-zero long-time limits $F_{\alpha\beta}(q;\infty)$ substituted at
the right-hand-side. 

\begin{figure}
\includegraphics[width=3in]{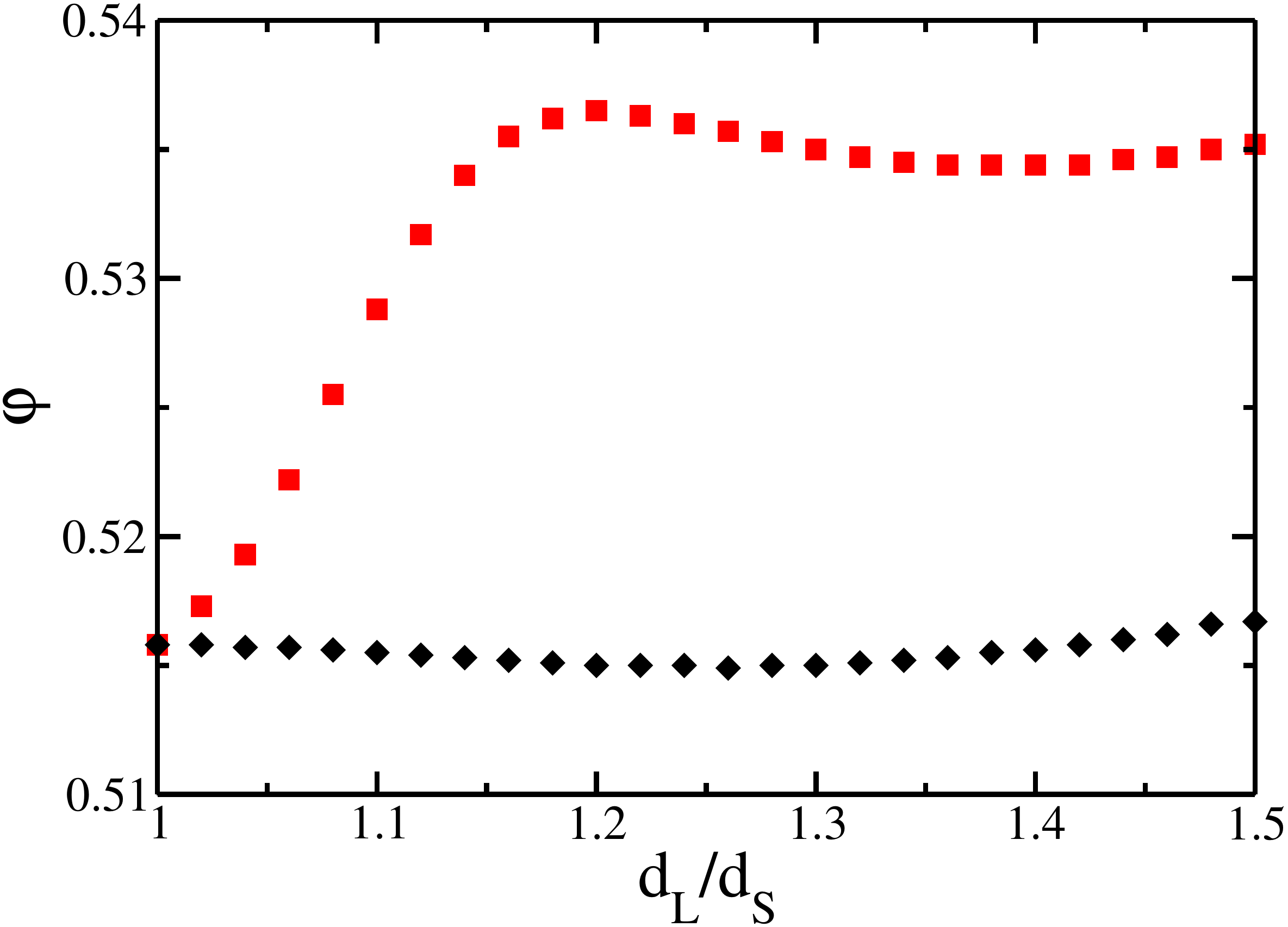}
\caption{\label{phased} 
Volume fraction $\varphi$ at the dynamic glass transition as a function
of the ratio of the hard sphere diameters $d_L/d_S$, 
for an equimolar binary hard-sphere mixture.
Red squares and black diamonds denote the location of the 
transition with and without particle size swaps.}
\end{figure}

We solved self-consistent equations (\ref{dgt}) for an equimolar binary hard-sphere
mixture using as the static input approximate equilibrium correlation functions
obtained from the Percus-Yevick closure \cite{PY1,PY2,Voigtmannthesis}. 
As shown in Fig. \ref{phased}, we found that for a system with particle size swaps 
the volume fraction at the dynamic glass transition increases with increasing 
particle diameter ratio up to the ratio of about 1.2, where the relative increase
is about 4\%, and then saturates. In agreement with Ref. \cite{GoetzeVoigtmann}, 
the location the dynamic glass 
transition for an equimolar mixture without swaps depends very weakly on
the ratio of particle diameters for the ratio smaller than 1.5.
In contrast, Ikeda \textit{et al.} \cite{IkedaZamponiIkeda} found that 
the volume fraction at the dynamic transition both without and with particle size 
swaps increases with the ratio of particle diameters; the absolute separation 
of these transitions increases monotonically but the relative separation saturates, 
although at a diameter ratio larger than found here \cite{FrancescoPC}.
The relative difference at the diameter ratio of 1.2 is approximately the same
according to both approaches. We note that
the MK model considered by Ikeda \textit{et al.} lacks non-trivial local 
structure but it is not clear whether this is the origin of the difference between 
our results and theirs. Finally, we show in the Supplemental Material that 
neglecting the coupling between ``normal'' dynamics and particle size swaps 
results in a phase diagram qualitatively similar to that showed in Fig. \ref{phased}.

\textit{Discussion. --} Wyart and Cates \cite{WyartCates,commentpar} have argued that the 
success of swap dynamics algorithms in equilibrating model systems at temperatures 
comparable to the laboratory glass transition temperature implies that the RFOT 
scenario needs to be re-evaluated. They conjectured that the 
dominant barriers for low temperature relaxation are local and the growing static 
correlation length is responsible for only a small fraction of slowing down. 
In contrast, Ikeda \textit{et al.} \cite{IkedaZamponiIkeda} 
argued that while the RFOT framework is still
valid, there is a need for a more general way to calculate the configurational entropy.
Effectively, they advocated using a less restrictive constrained equilibrium
construction in the derivation of the replica theory for systems with 
particle size swaps. This is equivalent to their more general \textit{Ansatz} for 
inter-replica correlations. 

In our approach, the basic functions that signal the dynamic glass transition
are the same for both ``normal'' and swap dynamics. Thus, we cannot account for the 
presence of particle size swaps by using a different \textit{Ansatz} for our
non-ergodicity parameters, $F_{\alpha\beta}(q;\infty)$. On the other hand, allowing
for particle size swaps does change the location of the dynamic transition and
influences the values of non-ergodicity parameters. We suggest that a possible way out
of this conundrum is to recognize the fact that metastable states, whose appearance
triggers the dynamic glass transition and which are counted 
by the configurational entropy, should be defined using a dynamical criterion. 
Some time ago \cite{SzamelEPL2010} we showed that the standard mode-coupling theory's
equation for the non-ergodicity parameter can be re-derived from a replica approach
combined with a dynamic criterion (vanishing of a current). It would be interesting
to check whether Eq. (\ref{dgt}) of the present theory can be re-derived
in a similar way \cite{commentcur}. 
We note that it is possible that different dynamics lead
to equivalent definitions of metastable states (\textit{e.g.} within mode-coupling
theory Newtonian and Brownian dynamics result in the same dynamic glass transition
scenario \cite{SL}). However, a significant modification
of the dynamics may result in different states being metastable and different
dynamic glass transition and configurational entropy scenarios. We leave these important
issues for future work.

\textit{Acknowledgments. --}
I thank the participants of the 2017 Royaumont meeting of the Simons Foundation 
``Cracking the Glass Problem'' collaboration for inspiring discussions, 
Th. Voigtmann for a reference to explicit expressions for binary hard 
sphere mixture PY structure factors and E. Flenner, and F. Zamponi for 
comments on the manuscript. 
I gratefully acknowledge partial support of NSF Grant No.~DMR-1608086.

\newpage
\setcounter{section}{1}
\setcounter{equation}{0}
\numberwithin{equation}{section}
\widetext

\widetext

\section*{Supplemental Material}

\subsection{I. \textit{The general theory}}

In this section we provide some additional information on the derivation
of the general structure of the theory.\\

First, let us outline the derivation of Eq. (8) of the main text.
On the one hand, the Laplace transform of the time derivative of $F_{\alpha\beta}(q;t)$
is equal to $zF_{\alpha\beta}(q;z)-F_{\alpha\beta}(q;t=0)\equiv
zF_{\alpha\beta}(q;z)-S_{\alpha\beta}(q)$. On the other hand, the same 
Laplace transform of the time derivative of $F_{\alpha\beta}(q;t)$ can be
re-written in the following way,
\begin{eqnarray}\label{Suppliden1}
&& \left<n_\alpha(\mathbf{q}) \Omega_\text{sw} 
\frac{1}{z-\Omega_\text{sw}} n_\beta(-\mathbf{q}) \right> 
= 
\left<n_\alpha(\mathbf{q}) \Omega_\text{sw} \mathcal{P} 
\frac{1}{z-\Omega_\text{sw}} n_\beta(-\mathbf{q}) \right>
+ \left<n_\alpha(\mathbf{q}) \Omega_\text{sw} \mathcal{Q} 
\frac{1}{z-\Omega_\text{sw}} n_\beta(-\mathbf{q}) \right> 
\nonumber \\ &=& 
\sum_{\gamma,\delta} 
\left<n_\alpha(\mathbf{q}) \Omega_\text{sw} n_\gamma(-\mathbf{q})\right>
\left<n_\gamma(\mathbf{q})n_\delta(-\mathbf{q})\right>^{-1}
\left< n_\delta(\mathbf{q}) \frac{1}{z-\Omega_\text{sw}} n_\beta(-\mathbf{q}) \right> 
\nonumber \\ && + 
\sum_{\gamma,\delta} \left<n_\alpha(\mathbf{q}) \Omega_\text{sw} \mathcal{Q} 
\frac{1}{z - \Omega_\text{sw}\mathcal{Q}} 
\mathcal{Q} \Omega_\text{sw} n_\gamma(-\mathbf{q})\right>
\left<n_\gamma(\mathbf{q})n_\delta(-\mathbf{q})\right>^{-1}
\left< n_\delta(\mathbf{q})\frac{1}{z-\Omega_\text{sw}}
n_\beta(-\mathbf{q}) \right>.
\end{eqnarray}
Here, we first inserted $\mathcal{I} \equiv \mathcal{P}+\mathcal{Q}$ between 
$ \Omega_\text{sw}$ and $\frac{1}{z-\Omega_\text{sw}}$ and then used the 
following exact identity
\begin{eqnarray}\label{Suppliden2}
\frac{1}{z-\Omega_\text{sw}} = \frac{1}{z - \Omega_\text{sw}\mathcal{Q}} +
\frac{1}{z - \Omega_\text{sw}\mathcal{Q}}\Omega_\text{sw}\mathcal{P} 
\frac{1}{z-\Omega_\text{sw}}.
\end{eqnarray}
Next, we define the frequency matrix, 
\begin{eqnarray}\label{SupplO22}
O_{\alpha\beta}(q) 
= -\left<n_\alpha(\mathbf{q})\Omega_{\text{sw}}n_\beta(-\mathbf{q})\right>
\end{eqnarray} 
and the matrix of reducible memory functions,
\begin{eqnarray}\label{SupplMred22}
M_{\alpha\gamma}^{\text{red}}(q;z) = 
 \left<n_\alpha(\mathbf{q}) \Omega_\text{sw} \mathcal{Q} 
\frac{1}{z - \Omega_\text{sw}\mathcal{Q}} 
\mathcal{Q} \Omega_\text{sw} n_\gamma(-\mathbf{q})\right>.
\end{eqnarray}
Using these definitions in Eq. (\ref{Suppliden1}) we can obtain 
Eq. (8) of the main text.\\

Second, we analyze matrices $O$ and $M^\text{red}$. We start with the frequency matrix,
\begin{eqnarray}\label{SupplO22str}
O_{\alpha\beta}(q) =
D_0 \mathbf{q}^2 x_\alpha \delta_{\alpha\beta} + 
\tau_\text{sw}^{-1}\left<n_\alpha(\mathbf{q})
\sum_i \left(1-S_i\right) w_i
n_\beta(-\mathbf{q})\right> = 
D_0 \mathbf{q}^2 x_\alpha \delta_{\alpha\beta}
+\frac{1}{2N\tau_\text{sw}}\left(2 \delta_{\alpha\beta} -1 \right) 
\left< \sum_i w_i \right>
\end{eqnarray}
and we note that it can be expressed in terms of 3-dimensional vectors 
$\mathsf{v}_\alpha$, $\alpha=A,B$, 
and a 3x3 matrix $\mathsf{O}$,
\begin{eqnarray}\label{SupplO33}
O_{\alpha\beta}(q) &=& \mathsf{v}_{\alpha}^{\text{T}}
\mathsf{O}(q) \mathsf{v}_{\beta}
\end{eqnarray}
Here $\mathsf{v}_{A}^{\text{T}}=(1,0,1)$, $\mathsf{v}_{B}^{\text{T}}=(0,1,-1)$, 
$\mathsf{O}_{11}=D_0q^2 x_A$, $\mathsf{O}_{22}=D_0q^2 x_B$, 
$\mathsf{O}_{33}=(1/2N\tau_{\text{sw}})\left<\sum_i w_i\right>$ and 
$\mathsf{O}_{ab}=0$ for $a\neq b$. This representation is convenient because 
it separates ``normal'' (\textit{i.e.} Brownian) dynamics represented by 
elements of matrix $\mathsf{O}$ 
with $a,b \le 2$ and the particle swaps represented by the 33 element
of $\mathsf{O}$. Next, we note the structure of the ``vertexes'' in the 
matrix of reducible memory functions,
\begin{eqnarray}\label{SupplM22vert}
\left. \mathcal{Q} \Omega_\text{sw} n_\alpha(-\mathbf{q})\right> &=& 
\left. \frac{D_0}{\sqrt{N}}
\mathcal{Q}\sum_i i\mathbf{q}\cdot\left( i\mathbf{q} - \beta\mathbf{F}_i\right)
\frac{1+\sigma_i\left(\delta_{\alpha A}-\delta_{\alpha B}\right)}{2}
e^{-i\mathbf{q}\cdot\mathbf{r}_i}\right>
- \left. \frac{1}{\sqrt{N}\tau_{\text{sw}}} \mathcal{Q} 
\sum_i w_i \sigma_i e^{-i\mathbf{q}\cdot\mathbf{r}_i}.
\right>\left(\delta_{\alpha A}-\delta_{\alpha B}\right)
\nonumber \\ &=& \sum_a \left. \pi_a(-\mathbf{q})\right> \mathsf{v}_{\alpha a},
\end{eqnarray}
where functions $\pi_a$ are defined in Eqs. (12-13) of the main
text and $\mathsf{v}_{\alpha a}$ is the $a$th element of vector $\mathsf{v}_{\alpha}$.
Eq. (\ref{SupplM22vert}) allows us to express matrix of reducible
memory functions $M^\text{red}$ in terms of  3-dimensional vectors 
$\mathsf{v}_\alpha$, $\alpha=A,B$, 
and a 3x3 matrix $\mathsf{M}^\text{red}$ given by Eq. (11) of the main text,
\begin{eqnarray}\label{SupplM33}
M_{\alpha\beta}^{\text{red}}(q;z) 
&=& \mathsf{v}_{\alpha}^{\text{T}}\mathsf{M}^{\text{red}}(q;z) 
\mathsf{v}_{\beta}.
\end{eqnarray}
Again, this representation allows us  
to separate ``normal'' (\textit{i.e.} Brownian) dynamics represented by 
elements of matrix $\mathsf{M}^\text{red}$ with $a,b \le 2$ 
and the particle swaps represented by the 33 element
of $\mathsf{M}^\text{red}$. We note, however, that in general there will be 
terms describing the time-delayed coupling between these relaxation channels. 
These coupling are described by elements $\mathsf{M}^{\text{red}}$ with
$a \le 2$ and $b=3$, and $a=3$ and $b\le 2$.\\

Third, we comment on the irreducible evolution operator and the relation
between matrices $\mathsf{M}^\text{red}$ and $\mathsf{M}^\text{irr}$. 
Cichocki and Hess [B. Cichocki and W. Hess, Physica A \textbf{141}, 475 (1987)]
realized that for systems with Brownian dynamics reducible memory functions 
can be reduced further by introducing the so-called irreducible evolution operator.
Later, Kawasaki  [K. Kawasaki, Physica A \textbf{215}, 61 (1995)] generalized this observation to systems with
stochastic dynamics, in particular to systems with dynamics described by a 
master equation.  
For systems evolving with Brownian dynamics without particle size swaps this
additional step has been given a clear diagrammatic interpretation [G. Szamel, 
J. Chem. Phys. \textbf{127}, 084515 (2007)].

For our present system, evolution operator $\Omega_\text{sw}$ consists of 
two parts, describing Brownian motion of particles, Eq. (2) and the
master equation-like part describing particle size swaps, Eq. (3),
respectively. 
Each of these parts leads to one-particle-reducible contributions. In order to 
remove these contributions, we introduced the following irreducible 
evolution operator,
\begin{eqnarray}\label{SupplOmegaswirr}
\Omega_{\text{sw}}^{\text{irr}} &=& \mathcal{Q}\Omega_{\text{sw}}\mathcal{Q} 
- \sum_{\alpha\beta} \left. \mathcal{Q} \Omega n_\alpha(-\mathbf{q})\right>
\left<n_\alpha(\mathbf{q})\Omega n_\beta(-\mathbf{q})\right>^{-1}
\left<n_\beta(\mathbf{q})\Omega
\mathcal{Q}\right.  
\\ \nonumber 
&& - \left. \mathcal{Q} \delta\Omega_{\text{sw}}\sigma(-\mathbf{q})\right>
\left<\sigma(\mathbf{q})\delta\Omega_{\text{sw}}\sigma(-\mathbf{q})\right>^{-1}
\left<\sigma(\mathbf{q})\delta\Omega_{\text{sw}}
\mathcal{Q}\right. ,
\end{eqnarray}
where $\sigma(\mathbf{q})$ is the microscopic composition field defined in
Eq. (15) of the main text. We emphasize that while there is some freedom 
in the definition of the irreducible evolution operator, the general principle is clear: 
one needs to remove one-particle-reducible parts from the projected evolution operator 
$\mathcal{Q}\Omega_{\text{sw}}\mathcal{Q}$. As noted in the main text, 
according to definition (\ref{SupplOmegaswirr}) reducible parts of 
$\mathcal{Q}\Omega_{\text{sw}}\mathcal{Q}$ are removed separately for the 
two relaxation channels in our system. We believe that this procedure can be 
given a diagrammatic interpretation, along the lines of Ref. [G. Szamel, 
J. Chem. Phys. \textbf{127}, 084515 (2007)].

Finally, we use the following identity, which is analogous to Eq. (\ref{Suppliden2}),
\begin{eqnarray}\label{Suppliden3}
\frac{1}{z-\mathcal{Q}\Omega_\text{sw}\mathcal{Q}} 
&=& \frac{1}{z - \Omega_\text{sw}^{\text{irr}}} +  
\frac{1}{z - \Omega_\text{sw}^{\text{irr}}}
\sum_{\alpha\beta} \left.  \mathcal{Q} \Omega n_\alpha(-\mathbf{q})\right>
\left<n_\alpha(\mathbf{q})\Omega n_\beta(-\mathbf{q})\right>^{-1}
\left<n_\beta(\mathbf{q})\Omega\mathcal{Q} \right.
\frac{1}{z-\mathcal{Q}\Omega_\text{sw}\mathcal{Q}}
\nonumber \\ && 
+ \frac{1}{z - \Omega_\text{sw}^{\text{irr}}}
\mathcal{Q}\left. \delta\Omega_\text{sw}\sigma(-\mathbf{q})\right>
\left<\sigma(\mathbf{q})\delta\Omega_\text{sw}\sigma(-\mathbf{q})\right>^{-1}
\left<\sigma(\mathbf{q})\delta\Omega_\text{sw} \right.
\mathcal{Q} 
\frac{1}{z-\mathcal{Q}\Omega_\text{sw}\mathcal{Q}},
\end{eqnarray}
to derive relation (16) between matrices $\mathsf{M}^{\text{red}}$ 
and $\mathsf{M}^{\text{irr}}$.\\

Fourth, we briefly comment on the structure of Eq. (8) of the main text 
in an approximation that neglects the time-delayed couplings between ``normal'' 
 (\textit{i.e.} Brownian) dynamics and particle size swaps. 
In this case, matrix $\mathsf{M}^\text{irr}$ has a block-diagonal structure, with
a 2x2 block that has the same form as that derived in the standard projection
operator analysis of a binary mixture [W. G\"otze and Th. Voigtmann, 
Phys. Rev. E \textbf{67}, 021502 (2003)] and a 1x1 block describing
the influence of the particle size swaps. The block-diagonal structure of
$\mathsf{M}^\text{irr}$ together with Eq. (16) of the main text implies
that matrix $\mathsf{M}^\text{red}$ also has a block-diagonal structure. This 
structure leads to two independent terms at the right-hand-side of Eq. (8).
After transforming the resulting equation back into the time domain we get
\begin{eqnarray}\label{SupplFMred}
\partial_t F_{\alpha\beta}(q;t) &=&  - \sum_{\gamma\delta} 
\int_0^t dt' \left(D_0 q^2 x_\alpha\delta_{\alpha\gamma}\delta(t-t') 
- M_{\alpha\gamma}^{\text{red,B}}(q;t-t')\right)
S^{-1}_{\gamma\delta}(q)
F_{\delta\beta}(q;t')
\nonumber \\ && 
- \sum_{\gamma\delta} 
\int_0^t dt' 
\left(\frac{1}{2N\tau_\text{sw}}\left<\sum_i w_i\right>\delta(t-t')  
- M^{\text{red,sw}}(q;t-t')\right)
\left(2\delta_{\alpha\gamma}-1\right)S^{-1}_{\gamma\delta}(q)
F_{\delta\beta}(q;t').
\end{eqnarray}
According to Eq. (\ref{SupplFMred}), the time-dependence of the 
density correlation functions is due to the presence of two independent relaxation
channels. The first channel describes the relaxation due the ``normal'' dynamics. 
The second channel describes the relaxation due to the particle 
size swaps. Each term involves its own reducible memory term, $M^{\text{red,B}}$
and $ M^{\text{red,sw}}$, respectively. It should be noted 
that the two relaxation channels act in parallel, rather than sequentially. 
This contrast with an implicit assumption made in Ref. [M. Wyart and M.E. Cates, 
Phys. Rev. Lett. \textbf{119}, 195501 (2017)]. We should emphasize that, 
in spite of the parallel arrangement of the two relaxation channels, the 
presence of the ``normal'' dynamics is essential since particle size swaps alone
cannot equilibrate the binary mixture.

\setcounter{section}{2}
\setcounter{equation}{0}
\subsection{II. \textit{Mode-coupling-like approximation}}

In this section we provide some additional information on the derivation of the
approximate expression for matrix $\mathsf{M}^\text{irr}$, Eq. (18)
of the main text.\\

The mode-coupling approximation is a sequence of three steps of somewhat 
different nature [W. G\"otze, \textit{Complex dynamics of glass-forming
liquids: A mode-coupling theory} (Oxford University Press, Oxford, 2008);
G. Szamel and H. L\"{o}wen, Phys. Rev. A \textbf{44}, 8215 (1991)]. 
First, the vertexes 
$\pi_a$, are projected on the subspace spanned by the parts of density
products orthogonal to the one-particle densities,
\begin{eqnarray}\label{Supplpi12proj}
\left. \pi_{1,2}(-\mathbf{q})\right> \!\! &\approx & \!\!
\frac{1}{4}\sum_{\epsilon, ...,\theta}\sum_{\mathbf{k}_1,...,\mathbf{k}_4}
\left. \mathcal{Q}n_\epsilon(-\mathbf{k}_1) n_\zeta(-\mathbf{k}_2)\right>
\left<n_\epsilon(\mathbf{k}_1) n_\zeta(\mathbf{k}_2) \mathcal{Q}
n_\eta(-\mathbf{k}_3) n_\theta(-\mathbf{k}_4)\right>^{-1}
\left<n_\eta(\mathbf{k}_3) n_\theta(\mathbf{k}_4) \pi_{1,2}(-\mathbf{q})\right>
\\ \label{Supplpi3proj}
\left. \pi_3(-\mathbf{q})\right>  \!\! &\approx & \!\! 
\frac{1}{4}\sum_{\epsilon, ...,\theta}\sum_{\mathbf{k}_1,...,\mathbf{k}_4}
\left. \mathcal{Q}n_\epsilon(-\mathbf{k}_1) n_\zeta(-\mathbf{k}_2)\right>
\left<n_\epsilon(\mathbf{k}_1) n_\zeta(\mathbf{k}_2) \mathcal{Q}
n_\eta(-\mathbf{k}_3) n_\theta(-\mathbf{k}_4)\right>^{-1}
\left<n_\eta(\mathbf{k}_3) n_\theta(\mathbf{k}_4) \pi_3(-\mathbf{q})\right>
\end{eqnarray}
Note that since the orthogonal projection is present in the definition
of functions $\pi_a$, we do not need to include it in the rightmost averages above,
\textit{i.e.} in 
$\left<n_\eta(\mathbf{k}_3) n_\theta(\mathbf{k}_4) \pi_{1,2}(-\mathbf{q})\right>$
and $\left<n_\eta(\mathbf{k}_3) n_\theta(\mathbf{k}_4) \pi_3(-\mathbf{q})\right>$. 
For systems with pairwise-additive interactions the projection in
Eq. (\ref{Supplpi12proj}) is just an exact transformation. In contrast,
due to non-pairwise-additive character of function $w_i$, 
the projection in Eq. (\ref{Supplpi12proj}) already introduces an approximation. 

Second, four-point dynamic correlation functions that appear after the
projections (\ref{Supplpi12proj}-\ref{Supplpi3proj}) are substituted into Eq. 
(17) of the main text are factorized and \textit{at the same time} 
the irreducible evolution operator is replaced by the original un-projected 
operator $\Omega_\text{sw}$,
\begin{eqnarray}\label{Suppl2ndapprdyn}
&& \left<n_\gamma(\mathbf{k}_1)n_\delta(\mathbf{k}_2)\mathcal{Q} 
\exp(\Omega_\text{sw}^{\text{irr}} t) 
\mathcal{Q} n_\epsilon(-\mathbf{k}_3)n_\zeta(-\mathbf{k}_4)\right>
\\ \nonumber &\approx&
\left<n_\gamma(\mathbf{k}_1) \exp(\Omega_\text{sw} t) n_\epsilon(-\mathbf{k}_3) \right>
\left<n_\delta(\mathbf{k}_2) \exp(\Omega_\text{sw} t) n_\zeta(-\mathbf{k}_4) \right>
+ \left<n_\gamma(\mathbf{k}_1) \exp(\Omega_\text{sw} t) n_\zeta(-\mathbf{k}_4) \right>
\left<n_\delta(\mathbf{k}_2) \exp(\Omega_\text{sw} t) n_\epsilon(-\mathbf{k}_3) \right>.
\end{eqnarray}
Eq. (\ref{Suppl2ndapprdyn}) is the major approximation of any mode-coupling-like
theory. Consistently, we also factorize the inverse matrix of the four-point 
static correlations,
\begin{eqnarray}\label{Suppl2ndapprstat}
&& \left<n_\epsilon(\mathbf{k}_1)n_\zeta(\mathbf{k}_2)\mathcal{Q} 
n_\eta(-\mathbf{k}_3)n_\theta(-\mathbf{k}_4)\right>^{-1}
\\ \nonumber &\approx&
\left<n_\epsilon(\mathbf{k}_1) n_\eta(-\mathbf{k}_3) \right>^{-1}
\left<n_\zeta(\mathbf{k}_2) n_\theta(-\mathbf{k}_4) \right>^{-1}
+ \left<n_\epsilon(\mathbf{k}_1) n_\theta(-\mathbf{k}_4) \right>^{-1}
\left<n_\zeta(\mathbf{k}_2) n_\eta(-\mathbf{k}_3) \right>^{-1}.
\end{eqnarray}

Third, we use additional approximations to calculate the remaining averages
in expressions (\ref{Supplpi12proj}-\ref{Supplpi3proj}), \textit{i.e.}
$\left<n_\eta(\mathbf{k}_3) n_\theta(\mathbf{k}_4) \pi_{1,2}(-\mathbf{q})\right>$
and $\left<n_\eta(\mathbf{k}_3) n_\theta(\mathbf{k}_4) \pi_3(-\mathbf{q})\right>$. 
The former calculation proceeds in the same way as that performed in the standard
mode-coupling theory for binary mixtures [W. G\"otze, \textit{Complex dynamics of glass-forming
liquids: A mode-coupling theory} (Oxford University Press, Oxford, 2008)]. Specifically, we 
use a mixture version of the convolution approximation and arrive at
\begin{eqnarray}\label{Supplvertex1}
\left< n_\eta(\mathbf{k}_3)n_\theta(\mathbf{k}_4)\pi_{1}(-\mathbf{q})
\right> = 
\delta_{\mathbf{k}_3+\mathbf{k}_4,\mathbf{q}} \frac{n}{\sqrt{N}} 
\sum_{\mu\nu} S_{\eta\nu}(k_3)
S_{\theta\mu}(k_4)\left[
\mathbf{k}_3 c_{\nu A}(k_3)\delta_{\mu A} + 
\mathbf{k}_4 c_{\mu A}(k_4)\delta_{\nu A}\right]\cdot \mathbf{q},
\end{eqnarray}
with the the other vertex,  
$\left< n_\eta(\mathbf{k}_1)n_\theta(\mathbf{k}_2)\pi_{2}(-\mathbf{q})\right>$,
being given by an analogous expression with $A$ replaced by $B$ at the 
right-hand-side.  As we mentioned in the main text, these two
vertexes have the same form as in the standard mode-coupling theory for 
binary mixtures [W. G\"otze, \textit{Complex dynamics of glass-forming
liquids: A mode-coupling theory} (Oxford University Press, Oxford, 2008)]. 
The calculation of 
$\left<n_\gamma(\mathbf{k}_3) n_\delta(\mathbf{k}_4) \pi_3(-\mathbf{q})\right>$
is a bit more involved. For an equimolar binary mixture, 
using the mixture version of the convolution 
approximation we initially obtain the following expression
\begin{eqnarray}\label{Supplvertex2a}
\left<n_\eta(\mathbf{k}_3) n_\theta(\mathbf{k}_4) \pi_3(-\mathbf{q})\right> &=& 
-\frac{\delta_{\mathbf{k}_3+\mathbf{k}_4,\mathbf{q}}}{\sqrt{N}\tau_\text{sw}}
\left[\frac{1}{2}
\left[R_\eta(k_3)\left(\delta_{\theta A}-\delta_{\theta B}\right)
+R_\theta(k_4)\left(\delta_{\eta A}-\delta_{\eta B}\right)
-Q\left(\delta_{\theta A}-\delta_{\theta B}\right)
\left(\delta_{\eta A}-\delta_{\eta B}\right)\right]
\right. \nonumber \\ && \left.  
-\frac{2}{N}\left<  \sum_{l=1}^{N}w_l \right>
\left(S_{\eta A}(k_3)S_{\theta A}(k_4) 
- S_{\eta B}(k_3)S_{\theta B}(k_4)\right)\right]
\end{eqnarray}
where 
\begin{eqnarray}\label{SupplRQa}
R_\theta(q) = N^{-1/2}\left< n_{\theta}(\mathbf{q})
\sum_{l=1}^{N}w_l e^{i\mathbf{q}\cdot\mathbf{r}_l}\right> 
\text{ and } Q = N^{-1}\left< \sum_{l=1}^N \sigma_l w_l \right>
\end{eqnarray}
To simplify expression (\ref{Supplvertex2a}) we factor out the average
$\left< \sum_l w_l\right>$ from $R_\theta$ and $Q$,
\begin{eqnarray}\label{SupplRQb}
R_\theta(q) \approx N^{-1}\left< \sum_{l=1}^{N}w_l \right>
\sum_{\lambda} S_{\theta\lambda}(q)  
\text{ and } Q \approx  
N^{-1}\left< \sum_{l=1}^{N}w_l \right> N^{-1}\left< \sum_{l=1}^N \sigma_l \right>= 0,
\end{eqnarray}
where in the last equality we used the equimolar condition 
$\left<\sum_{l=1}^N \sigma_l \right>=0$. Using Eq. (\ref{SupplRQb}) in 
Eq. (\ref{Supplvertex2a}) we obtain the expression given in Eq. (19) of
the main text,
\begin{eqnarray}\label{Supplvertex2b}
\left<n_\eta(\mathbf{k}_3) n_\theta(\mathbf{k}_4) \pi_3(-\mathbf{q})\right> &=& 
-\delta_{\mathbf{k}_3+\mathbf{k}_4,\mathbf{q}}
\frac{\left<  \sum_l w_l \right>}{2 N^{3/2}\tau_\text{sw}}\left[
\sum_{\lambda} S_{\eta\lambda}(k_3) \left(\delta_{\theta A}-\delta_{\theta B}\right)
+\sum_{\lambda} S_{\theta\lambda}(k_4) \left(\delta_{\eta A}-\delta_{\eta B}\right)
\right. \\ \nonumber && \left.  
-4
\left(S_{\eta A}(k_3)S_{\theta A}(k_4) 
- S_{\eta B}(k_3)S_{\theta B}(k_4)\right)\right]
\\ \nonumber  &=&
- \delta_{\mathbf{k}_3+\mathbf{k}_4,\mathbf{q}}
\frac{\left< \sum_l w_l \right>}{2 N^{3/2}\tau_{\text{sw}}}\sum_{\mu}
\left[\sum_\lambda \left(
S_{\eta\lambda}(k_3)\delta_{\theta\mu} + S_{\theta\lambda}(k_4)\delta_{\eta\mu}\right)
-4 S_{\eta\mu}(k_3)S_{\theta\mu}(k_4)\right]
\left(\delta_{\mu A}-\delta_{\mu B}\right)
\end{eqnarray}

\setcounter{section}{3}
\setcounter{equation}{0}
\subsection{III. \textit{Dynamic glass transition phase diagram without
coupling between two relaxation channels}}

\begin{figure}
\includegraphics[width=3in]{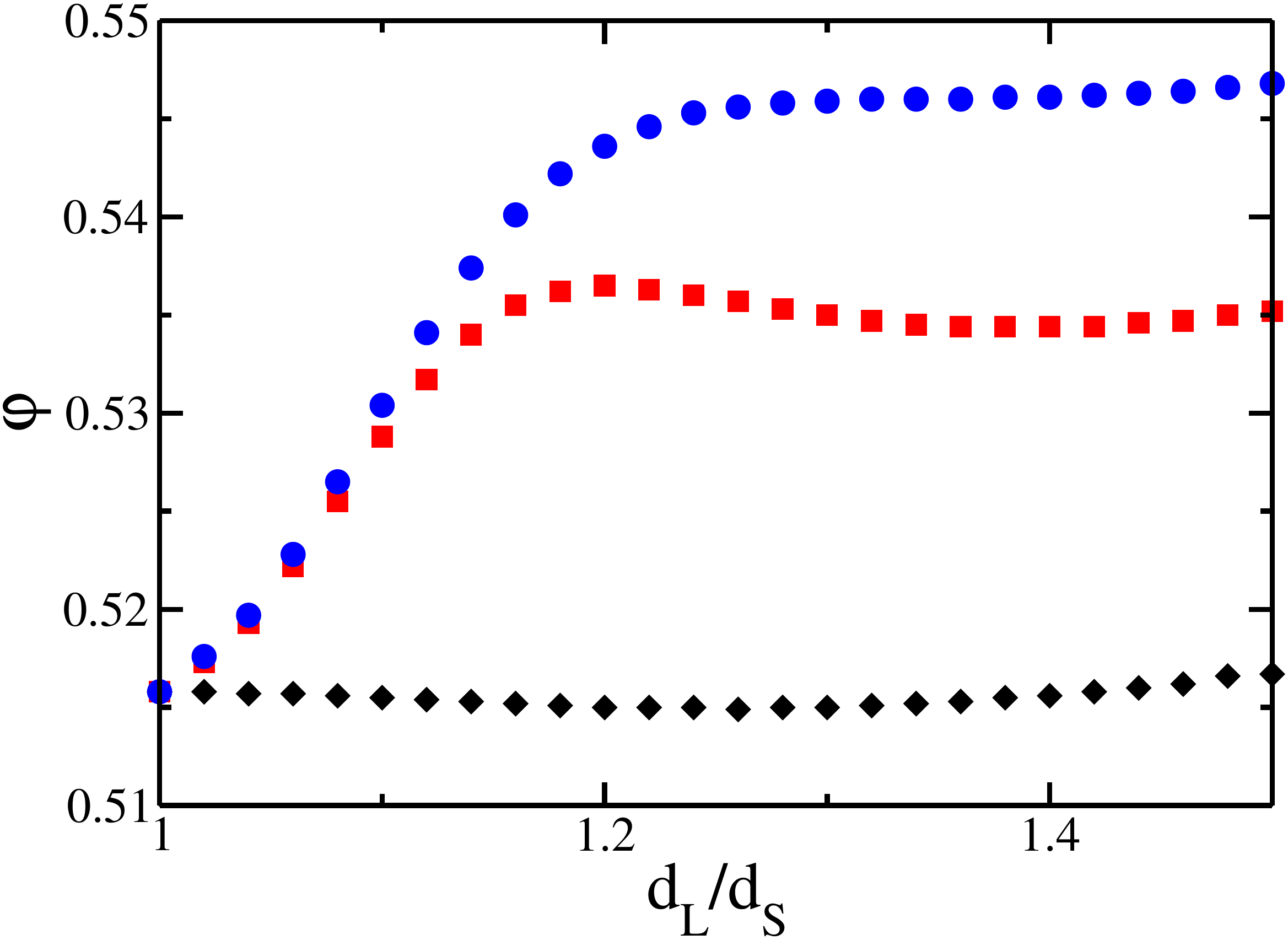}
\caption{\label{Supplphased} 
Volume fraction $\varphi$ at the dynamic glass transition as a function
of the ratio of the hard sphere diameters $d_L/d_S$, 
for an equimolar binary hard-sphere mixture.
Red squares and black diamonds denote the location of the 
transition with and without particle size swaps. Blue circles denote
the location of the transition with particle size swaps but without 
dynamic coupling between two relaxation channels.}
\end{figure}

We argued in the main text and at the end of Sec. I of the Supplemental Material that
relaxation channels associated with ``normal'' dynamics and particle size swaps act
approximately in parallel. To explore this issue farther we calculated the dynamic
glass transition phase diagram neglecting dynamic coupling between these
two relaxation channels, \textit{i.e,} neglecting elements of matrix 
$\mathsf{M}^{\text{irr}}(q;\infty)$ with $a=1,2$ and $b=3$, and $a=3$ and $b=1,2$.
As shown in Fig. \ref{Supplphased}, 
this additional approximation results in a qualitatively the
same phase diagram. This supports our claim that the two relaxation channels
act approximately in parallel. Interestingly, the coupling between the 
relaxation mechanisms diminishes the shift of the dynamic glass transition and
thus likely slows down the dynamics. 

\begin{thebibliography}{99}
\bibitem{comment} Continuous polydispersity allows one to treat the particle diameter 
as an additional continuous variable, with its own equation of motion 
\cite{BritoLernerWyart}. In the context of our model, this would lead to an additional
diffusive term in the equation of motion for the $N$-particle distribution, Eq.
(\ref{Peom}) rather than a master equation-like term $\delta \Omega_{\text{sw}}$,
Eq. (\ref{deltaOmegasw}).
\bibitem{commentpar} We note that Wyart and Cates assumed that allowing 
particle size swaps lowers one contribution to the effective activation energy. This
seems to imply that relaxation channel associated with particle 
size swaps is sequential rather than (approximately) parallel to the ``normal''
dynamics.
\bibitem{commentcur} This could be easier to do for a system with a continuous 
polydispersity since in this case one could define a generalized current with
components in both physical space and diameter space.
\end{thebibliography}
\end{document}